\documentclass[12pt]{article}
\usepackage{graphicx}
\usepackage{epsfig}
\usepackage[figuresright]{rotating}
\usepackage{amsmath}
\usepackage{amssymb}
\usepackage{amsfonts}
\usepackage{array}

\renewcommand{\thefootnote}{\fnsymbol{footnote}}

\usepackage{cite}
\usepackage{amssymb}
\usepackage{amsfonts}

\begin{document}

\begin{center}
{\LARGE{\bf Mass and structure of the nucleon: Gluon trace anomaly versus
    spontaneous symmetry breaking}}\\
[1ex] 
Martin Schumacher\footnote{mschuma3@gwdg.de}\\
II.  Physikalisches Institut der Universit\"at G\"ottingen,
Friedrich-Hund-Platz 1\\ D-37077 G\"ottingen, Germany\\

\end{center}

\begin{abstract}
Two different approaches to mass and structure of the nucleon are discussed
in recent works, 
{\it viz.} (case i) the QCD lagrangian evaluated via lattice calculations
and (case ii) spontaneous symmetry breaking mediated by
the $\sigma$ field. These approaches are complementary in the sense that the 
QCD lagrangian makes use of the  gluon content of the nucleon 
entering in terms of  the gluon trace-anomaly
and ignores the effects of $q\bar{q}$ vacuum polarization, whereas in
spontaneous symmetry breaking masses are formed by attaching $q\bar{q}$ 
pairs to the valence quarks, thus giving them a definite mass which is named
the constituent mass. By the same process the $q{\bar q}$ pairs of the vacuum
polarization acquire mass and in this form  are the elements of the quark
condensate, having an up-quark and a down-quark component. A linear combination
of these two components in the form $\sigma=1/\sqrt{2}(u \bar{u}+d\bar{d})$
shows up as the $\sigma$ field. It is shown that (case i) corresponds to an
unstable nucleon configuration whereas (case ii) corresponds to a stable 
nucleon configuration  as observed in low-energy 
photo-nuclear and pion-nuclear reactions.  
\end{abstract}

\renewcommand{\thefootnote}{\arabic{footnote}}

\section{Introduction}

In the last decades great effort has been devoted to the question
how far hadrons can be described in terms of first-principle QCD and
what effective forms of QCD are valid in the low-energy limit.
First-principle QCD makes use of the QCD lagrangian containing the 
gluon trace-anomaly and the masses and wave-functions  of the current
quarks. Constructing a nucleon in terms of these two components leads to a
nucleon mass in the form $M_N=M_0+\sigma_{\pi N}$, where $M_0$ is the gluonic
part 
remaining nonzero in the chiral limit and $\sigma_{\pi N}$ the $\sigma$ term
which 
vanishes in the chiral limit. The term ``chiral limit'' denotes the
hypothetical case where the effects of the Higgs boson are disregarded.
The $\sigma$ term is known from experimental
investigations and has a widely adopted value of $\sigma_{\pi N}=45$
MeV. Then, 
from a nucleon mass of $M_N=939$ MeV we arrive at $M_0=894$ MeV. Lattice QCD
is used to confirm that this latter quantity can be traced back to the
gluonic trace anomaly. The present status of these activities will be reviewed
in  the following sections.

The linear $\sigma$ model (L$\sigma$M)  and the Nambu--Jona-Lasinio 
(NJL)  model both are effective field theories 
for the mass of constituent quarks and of the $\sigma$ meson. These
two effective field theories  had the characteristics of a toy model
as long as the $\sigma$ meson had not been discovered. This has changed
completely after the $\sigma$ meson  has been definitely observed
as part of the constituent-quark structure in a Compton scattering
experiment where  a mass of this particle of $m_\sigma= 600\pm 70$ MeV has 
been  determined \cite{schumacher13}. Arguments based on the NJL 
model led to a mass of
$m_\sigma= 666$ MeV \cite{schumacher14,schumacher16}. 
This latter mass proved to be compatible with three
fundamental structure constants of the nucleon, {\it viz.} the mass, the 
magnetic
moment and the polarizability.  This means that as far as the
constituent-quark mass and structure and the $\sigma$-meson mass and structure
are concerned, the NJL model is the low-energy effective field theory of first
choice.

\section{Status of lattice calculations based on the gluon trace
 anomaly}

For the nucleon electroweak (EW) spontaneous symmetry breaking mediated by the
Higgs boson generates only 2\% of the observed mass. The missing 98\%
are explained in two different ways, {\it viz.} \\
$\bullet$  via perturbative QCD evaluated through calculation on the lattice
and\\ 
$\bullet$ via low-energy QCD as provided by the NJL model.

In  QCD the trace of the energy momentum tensor becomes 
\cite{donoghue92,thomas00    } 
\begin{equation}
\theta^\mu_\mu=\frac{\beta(g)}{2g}Tr(G_{\mu\nu} G^{\mu\nu})+\sum_{\rm flavors}
m_t\bar{\psi}_i\psi_i,
\label{trace}
\end{equation}
where $\beta(g)$ is the QCD $\beta$ function. Apart from the quark mass term 
the energy momentum tensor has an extra term proportional to the squared 
gluon field tensor, which is referred to as the trace anomaly. We ignore the
heavy quarks (using $N_f=3$) and work to leading order in $\alpha_s$. Then,
with the normalization $\bar{u}u=2 M_N$ the nucleon mass becomes
\begin{equation}
M_N=(2M_N)^{-1}( \langle N(p)|-\frac{9\alpha_s}{4\pi}Tr(G_{\mu\nu}G^{\mu\nu})
+ m_u\bar{\psi}_u\psi_u   + m_d\bar{\psi}_d\psi_d   + m_s\bar{\psi}_s\psi_s
|N(P)\rangle ),
\label{masstot}
\end{equation} 
where the trace anomaly term survives in the chiral limit. The trace anomaly
including the strange quark leads to 
\begin{equation}
M_0=(2M_N)^{-1} (\langle N(p)|-\frac{9\alpha_s}{4\pi}Tr(G_{\mu\nu}G^{\mu\nu})
  + m_s\bar{\psi}_s\psi_s
|N(P)\rangle )=894 \mp 8 \quad {\rm MeV},
\label{massglue}
\end{equation} 
and the term including the up and down quark
\begin{equation}
\sigma_{\pi N}=(2M_N)^{-1} (\langle N(p)|
+ m_u\bar{\psi}_u\psi_u   + m_d\bar{\psi}_d\psi_d  
|N(P)\rangle)=45\pm 8 \quad {\rm MeV}. 
\label{masssigma}
\end{equation} 
Eqs. (\ref{masstot}) and (\ref{massglue}) contain the effects of the strange
quark which, however, may be disregarded in case of the nucleon.

The mass value given in Eq. (\ref{masssigma}) has been obtained by experiments.
The mass term given in Eq. (\ref{massglue}) is the difference between the
nucleon mass and the mass given  in Eq. (\ref{masssigma}). Therefore,
the two errors are anticorrelated. QCD on the lattice is used to verify the mass
value of Eq. (\ref{massglue}). A problem entering into these lattice
calculations is that sufficiently small lattice constants lead to too large
computer times. Therefore, the  lattice calculations have to be carried 
out at unphysically large lattice constants. Thereafter, extrapolations to the
physics point have to be made. The tool for these extrapolations is borrowed
from  baryon chiral perturbation theory ($B\chi PT$).
The status of these lattice calculations is
described in \cite{alvares13}.   In \cite{alvares13} the nucleon mass $M_N$
and the $\sigma_{\pi}$ term are studied in the covariant baryon
perturbation theory ($B\chi PT$)  up to chiral order $p^4$. Fits have been
made using $B\chi PT $ with and without explicit $\Delta$ degrees of freedom to
combined lattice QCD (lQCD) data from various collaborations. In total
10 low-energy constants (LECs) were needed, some of which have been fitted to
the lQCD data. The masses obtained through these fits are typically
in the range 
$M_N = 862 - 904$ and the $\sigma$ terms in the range 
$\sigma_{\pi N}= 64 - 36$ MeV.   Further information is given in 
\cite{bali13,procura04}.

The disadvantage of representing the nucleon mass in terms of the gluon trace
anomaly is that this representation does not lead to a reasonable model
of the nucleon.  In \cite{donoghue92} it has been pointed out that a nucleon
with properties predicted by the QCD lagrangian does not go well with
the widely accepted constituent-quark model of the nucleon. The explanation
is that  constituent quarks  are a property of the nucleon in the low-energy
limit which may be investigated in low-energy photo-nuclear and pion-nuclear
reactions.
These reactions definitely show that the low-energy structure
of the nucleon does not indicate any sign of a gluon content, but rather
constituent quarks and mesons are observed. In case of Compton scattering 
the process
consists of two parts, named the $s$-channel part and the $t$-channel
part. The $s$-channel part proceeds through resonant and nonresonant
excitation of the nucleon, the $t$-channel part through a
$\sigma$ meson located on the constituent quarks.

\section{Illustration of the topic in terms of the mexican-hat potential}

\begin{figure}[h]
\begin{center}
\includegraphics[width=0.6\linewidth]{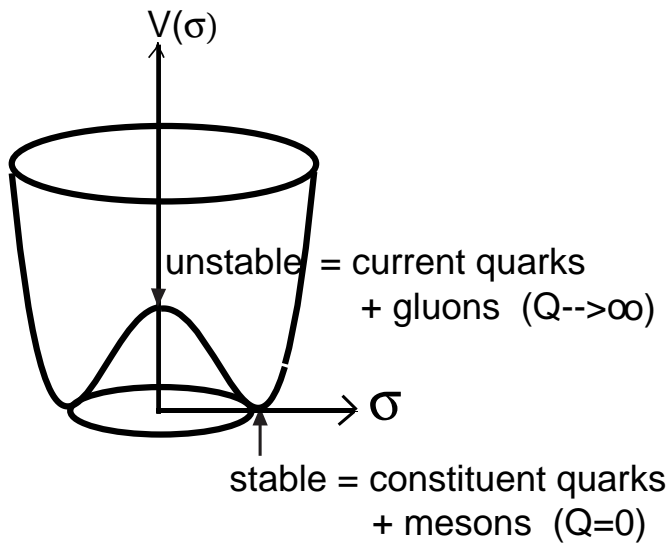}
\end{center}
\caption{Mexican hat potential for the $\sigma$ meson. In the center of
the potential we find the current-quark gluon phase. At the expectation value
of the $\sigma$ field we find the constituent-quark meson  phase.}
\label{sombr13}
\end{figure}

The structure of the nucleon appears different depending on the type of
investigation. The  parameter  discriminating between the different types of
observed structures is the momentum transfer $Q$ introduced in connection with
deep inelastic electron, muon and neutrino scattering experiments. Using 
Heisenberg's uncertainty principle in the form $Q \Delta x = \hbar$ we may 
interpret
$\Delta x=\hbar/Q$ as the spacial resolution of the experiment.
The range of structures extends
from asymptotic freedom $(Q\to \infty)$  to confinement $(Q=0)$. 
In these two limiting cases the structures of the nucleon and the degrees
of freedom are different, i.e. quarks and gluons in the 
$(Q\to \infty)$ case and constituent quarks and mesons in the 
$(Q=0)$ case. In the $(Q\to \infty)$ case the interquark coupling
constant is small $\alpha_s(Q\to \infty)\to 0$. This has the consequence
that the $q\bar{q}$ current-quark Dirac sea is completely decoupled 
from  the current-quarks and gluons of the valence sector. 
In Figure 1 this case corresponds to the center of the mexican-hat
potential. In terms of the mexican-hat potential this is an unstable equilibrium
corresponding to an unstable form  of the nucleon structure.  The appropriate
theoretical approach to this modification is perturbative QCD. There is a
widespread  belief that the mass of the nucleon may be given by
Eq. (\ref{masstot}), i. e. by a current-quark component and a gluon component.
The latter component requires some explanation. We know that in case of 
an  electron the related electric field does  not contribute to the 
mass of the electron whereas
in case of  gluons it is believed that the gluon component is responsible
for $\sim 95 \%$ of the nucleon mass. This difference between the
electromagnetic and the gluon case is related to  the fact that gluons
carry color charges leading to a gluon-gluon interaction. This gluon-gluon
interaction makes resonant gluonic states named  glue-balls possible.
This leads to the mass of the nucleon but not to a reasonable model
of the nucleon as observed in the low-energy domain.

Up to this point we have described the structure of the nucleon as
corresponding to the maximum of the mexican hat potential. Here the
relevant degrees of freedom are current quarks and gluons which cannot
provide a  reasonable model of the nucleon as observed experimentally in the
low-energy domain.
This is different when we use the tools of low-energy QCD, i.e. the tools
as provided by the NJL model. The NJL model provides us with a quantitative
description of the physics in the minimum  of the mexican hat
potential (see Figure \ref{sombr13}) and to realistic degrees of freedom in
the low-energy domain. These are constituent quarks and mesons.

The minimum of the mexican hat potential is
located at the vacuum expectation
$v=\langle \sigma \rangle$ value  of the $\sigma$ field. This vacuum 
expectation
value is given by the pion decay constant in the chiral limit (cl)
\begin{equation}
v\equiv \langle \sigma \rangle \equiv f_\pi^{\rm cl}= {\rm 89.8 \, MeV}.
\label{vacexp}
 \end{equation}
Then the mass of the $\sigma$ particle
in the chiral limit is given by \cite{schumacher14,schumacher16}
\begin{equation}
m^{\rm cl}_\sigma= 2M=\frac{4 \pi}{\sqrt{3}}f^{\rm cl}_\pi=652 \,{\rm MeV}.
\label{sigmamass}
\end{equation}
The mass of the constituent quark in the chiral limit M entering into
Eq. (\ref{sigmamass}) is related to the
pion decay constant $f^{\rm cl}_\pi$ in the chiral limit via the following
two relations \cite{schumacher16}
\begin{eqnarray}
&&M=-\frac{8 i N_c g^2}{(m^{\rm cl}_\sigma)^2}
\int\frac{d^4p}{(2\pi)^4}\frac{M}{p^2-M^2}, \label{conmass}\\
&&f^{\rm cl}_\pi=- 4 i N_c g M \int \frac{d^4 p}{(2 \pi)^4}\frac{1}{(p^2-M^2)^2}
\label{condecay}
\end{eqnarray}
leading to
\begin{equation}
M=\frac{2\pi}{\sqrt{3}} f^{\rm cl}_\pi.
\label{Mf}
\end{equation}

From the mass of the sigma meson  in the chiral limit as given
Eq. (\ref{sigmamass})
 the mass of the $\sigma$ meson with the effects
of the Higgs boson included is obtained via
\begin{equation}
m_\sigma = m^{\rm cl}_\sigma + m^0_u + m^0_d = 666\, {\rm MeV},
\label{sigmamass1}
\end{equation}
where $m^0_u= 5 \,{\rm MeV}$ and $m^0_d= 9 \, {\rm MeV}$ are the current-quark
masses of the up quark and the down quark, respectively. The same result
is obtained from the relation
\begin{equation}
m_\sigma=\sqrt{(m^{\rm cl}_\sigma)^2+ ({\hat m}_\pi)^2}=666 \,{\rm MeV},
\label{sigmamass2}
\end{equation}
where ${\hat m}_\pi$ is the average pion mass. Eq. (\ref{sigmamass2}) follows
from the NJL theory. Eq. (\ref{sigmamass1}) follows from (\ref{sigmamass2})
by applying the Gell-Mann-Oaks-Renner (GOR) relation.
For details see \cite{schumacher14,schumacher16}. The
constituent-quark masses with the effects of the Higgs boson included
are given by
\begin{eqnarray}
m_u=\frac12 m^{\rm cl}_\sigma +m^0_u=331\, {\rm MeV},\label{const-u}\\
m_d=\frac12 m^{\rm cl}_\sigma +m^0_d=335\, {\rm MeV}.\label{const-d}
\end{eqnarray}
The relations (\ref{const-u}) and  (\ref{const-d}) are valid because the
binding of the two constituent quarks in the $\sigma$ meson is small so that
the mass of the $\sigma$ may be regarded as the sum of the masses
of the two constituent quarks.

\section{The  mass and structure of the
 nucleon} 

In the foregoing section we have shown that the constituent-quark masses
$m_u=331$ MeV and $m_d=335$ MeV are well founded values in the
framework of chiral symmetry breaking. Differing from the case of the
$\sigma$ meson, in case of the nucleon we have to take into account the
effects of a binding energy which reduces the mass. Without this
mass reduction the mass of the nucleon would be
\begin{equation}
m^0_p = 997\,{\rm MeV},\quad m^0_n= 1001\, {\rm MeV}.
\label{nucleonmass1}
\end{equation}
This leads to a binding energy B/A per constituent quark (A=3) in the two
isospin partners $p$ and $n$ as given in the following table.
\begin{table}[h]
\caption{Binding energy B/A per constituent quark (A=3) in the two
isospin partners p and n.}
\begin{center}
\begin{tabular}{c|c|c}
\hline
nucleon& p & n\\
\hline
B/A & 19.6 MeV& 20.5 MeV\\
\hline
\end{tabular}
\end{center}
\label{table1}
\end{table}
Constituent quarks are the building blocks of the nucleon in a way closely 
resembling that of nucleons in a nucleus. A difference between the two cases
is that constituent quarks in a nucleon cannot be extracted from the nucleon
because of the
color charges. The binding energy  B/A of constituent quarks in a nucleon
is a consequence of the interquark forces which are mediated by mesons
as in case of nucleons in a nucleus. We, therefore expect binding energies
B/A per constituent quark to be of the same order of magnitude as the 
binding energies
B/A  per nucleon in the light nuclei $^3_1$H and $^3_2$He.
These are are 2.83 MeV and 2.57 MeV, respectively.
This means that the binding energies of constituent quarks in the nucleons
are about a factor 7.5 larger than the corresponding numbers for nucleons in
the light nuclei  $^3_1$H and $^3_2$He. This result is very plausible
because of the smaller distances in the nucleon and because 
of possible residual gluonic components in the inter-quark forces. 
These residual gluonic 
components are not expected in case of nucleons in nuclei.

In an approach to understand the difference of the binding energies
$B_p= 59$ MeV and $B_n=61$ MeV
for the proton  and the neutron, respectively, we write down the 
interquark Coulomb energy in
the form 
\begin{equation}
U=\sum_{i,j,i<j}\frac{e_ie_j}{r_{ij}}\alpha_{em}\hbar c.
\label{coulomb}
\end{equation} 
Then with a plausible adopted value of $\langle r_{ij} 
\rangle \approx 0.24$ fm we arrive at $U_p\approx 0$
MeV and $U_n\approx -2.0$ MeV. 
This means that the interquark electric force does not have an effect 
on the proton mass but reduces the neutron mass by 2.0 MeV.
This  consideration explains the
difference of the B values for the proton and the neutron as being 
due to a Coulomb attraction in the neutron. For details see \cite{schumacher16}.

\section{Summary and conclusion}
As a conclusion we may state
that the NJL model together with a small contribution due to Coulomb forces
quantitatively explains the mass of the nucleon at a percent level of
precision. It has been shown previously
\cite{schumacher14} that this approach also leads to a quantitative explanation
of the fundamental structure constants of the nucleon, {\rm viz.}
the magnetic moment and the electromagnetic polarizability.

This success contrasts with the current-quark-gluon approach where a
model of the nucleon is applied which does not contain the experimentally
observed structure of the nucleon at low energies. Furthermore, the 
computational tools
entering into the approach provided by lattice calculations
are only applicable for comparatively large 
lattice constants, so that extrapolations to the physical point are required.
The method making this extrapolation possible is provided by B$\chi$PT.


\begin{thebibliography}{99}
\bibitem{schumacher13}
Martin Schumacher and Michael D. Scadron, Fortschr. Phys. {\bf 61}, 703 (2013)
arXiv:1301.1567[hep-ph].


\bibitem{schumacher14}
Martin Schumacher, Ann.Phys. (Berlin) {\bf 526}, 215 (2014); arXiv: 
1403.7804 [hep-ph].

\bibitem{schumacher16}
Martin Schumacher, Pramana, - J. Phys.  87 : {\bf 44} (2016) ; arXiv:1506.00410
[hep-ph]. 


\bibitem{donoghue92}
John F. Donoghue, Eugene Golowich, Barry R. Holstein ``Dynamics of the
Standard Model'' Cambridge Monographs (1992).


\bibitem{thomas00}
Anthony W. Thomas, Wolfram Weise, ``The Structure of the Nucleon'' Wiley-VCH
(2000). 



\bibitem{alvares13}
L. Alvares-Ruso, T. Ledwig, J. Martin Camalich, and M.J. Vicente-Vacas,
Phys. Rev. D 88, 054507 (2013), arXiv:1304.0483 [hep-ph].

\bibitem{bali13}
G.S. Bali, et al., Nucl. Phys. B 866 (2013) 1- 25, arXiv:1206.7034 [hep-lat].

\bibitem{procura04}
M. Procura, T.R. Hemmert, W. Weise, Phys. Rev. D 69 (2004) 034505,
arXiv:hep-lat/0309020,   

\end{thebibliography}
\end{document}